\documentclass[journal,11pt,doublecolumn]{IEEEtran}
\usepackage[defernumbers=true, backend=biber, sorting=none, maxnames=4]{biblatex}
\usepackage[usenames,dvipsnames]{xcolor}
\usepackage{amsmath}
\usepackage{amsfonts}
\usepackage{hyperref}
\usepackage{graphicx}
\usepackage{authblk}
\usepackage{dsfont}
\usepackage{mathtools}
\usepackage{amssymb}
\usepackage{enumerate}
\usepackage{enumitem}
\usepackage{tabularx}
\usepackage{makecell}
\usepackage{settings_custom}
\usepackage{caption}
\newcommand{\new}[1]{#1}

\usepackage{float}

\makeatletter
\let\blx@rerun@biber\relax
\makeatother

\newcommand{\jdcomment}[1]{{\bf \color{magenta} JD: #1}}

\newtheorem{conjecture}{Conjecture}

\begin{document}

\title{Phase Retrieval: 
From Computational Imaging to Machine Learning}
\author[1]{Jonathan Dong}
\author[2]{Lorenzo Valzania}
\author[3]{Antoine Maillard}
\author[1]{Thanh-an Pham}
\author[2]{Sylvain Gigan}
\author[1]{Michael Unser}
\affil[1]{\small Biomedical Imaging Group, \'Ecole polytechnique fédérale de Lausanne (EPFL), Lausanne, Switzerland \normalsize}
\affil[2]{\small Laboratoire Kastler Brossel, \'Ecole Normale Sup\'erieure – Paris Sciences et Lettres (PSL) Research University, Sorbonne Université, CNRS UMR 8552, Collège de France, 24 rue Lhomond, Paris, France \normalsize} 
\affil[3]{\small Department of Mathematics \& Institute for Mathematical Research (FIM), ETH Zürich, Zürich, Switzerland \normalsize}
\date{}

\maketitle

\vspace{-1.3cm}
\begin{abstract}
    \new{Phase retrieval consists in the recovery of a complex-valued signal from intensity-only measurements. As it pervades a broad variety of applications, many researchers have striven to develop phase-retrieval algorithms. Classical approaches involve techniques as varied as generic gradient-descent routines or specialized spectral methods, to name a few. Yet, the phase-recovery problem remains a challenge to this day. Recently, however, advances in machine learning have revitalized the study of phase retrieval in two ways: significant theoretical advances have emerged from the analogy between phase retrieval and single-layer neural networks; practical breakthroughs have been obtained thanks to deep-learning regularization. In this tutorial, we review phase retrieval under a unifying framework that encompasses classical and machine-learning methods. We focus on three key elements: applications, overview of recent reconstruction algorithms, and the latest theoretical results.}
\end{abstract}
\vspace{-3mm}


\section{Introduction}

\begin{figure*}
    \centering
    \includegraphics[width=\linewidth]{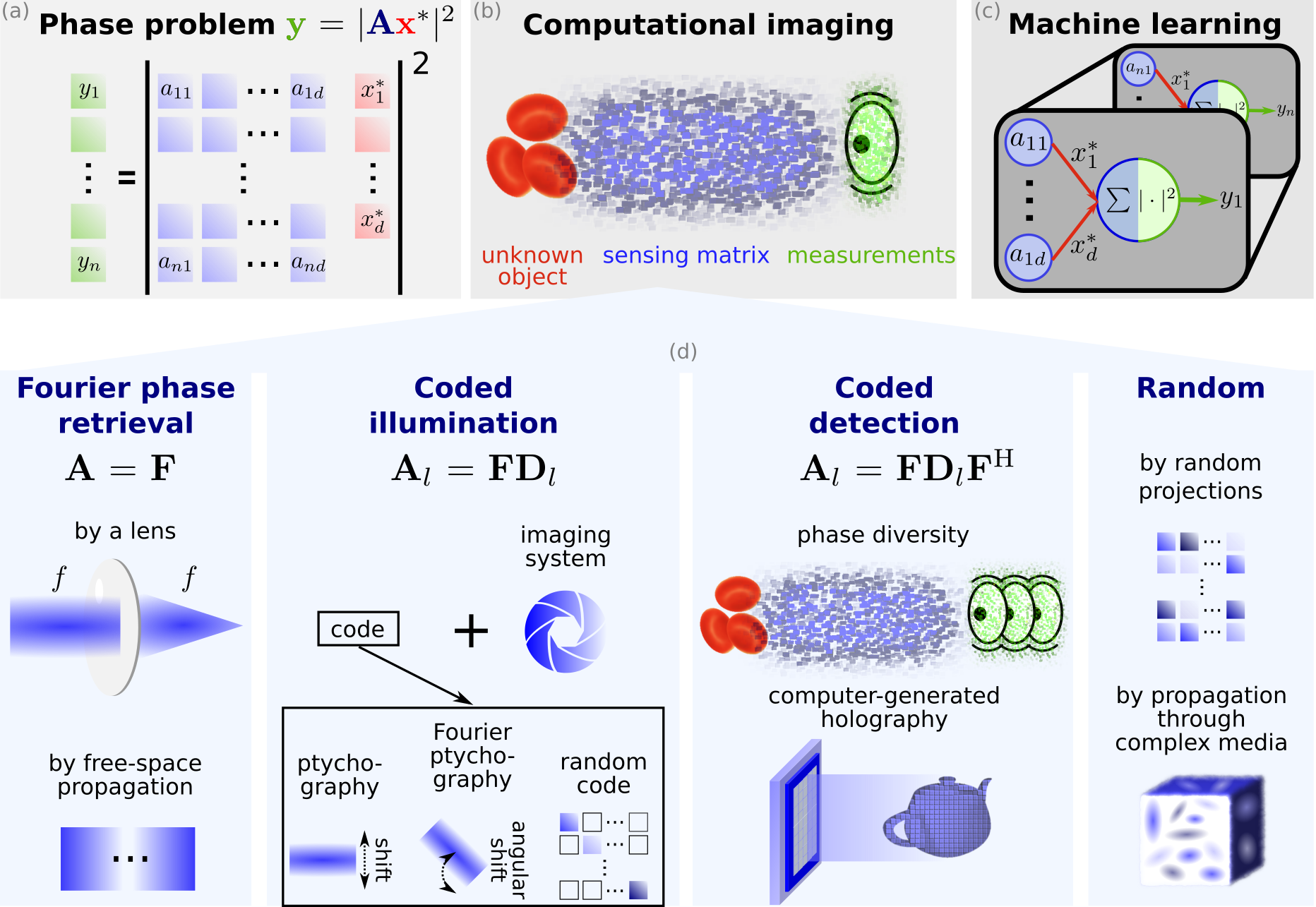}
    \caption{Unifying framework for phase-retrieval applications. The nonlinear phase-retrieval equation (a) is encountered in many computational imaging settings (b) and corresponds to single-layer neural-network optimization (c). Depending on the measurement matrix $\bA$, four different classes of phase-retrieval problems can be encountered (d) in imaging applications. 
    }
    \label{fig:drawing}
\end{figure*}

Phase retrieval is a longstanding computational problem that is simple to define yet difficult to solve. It consists in the $d$-dimensional search for $\bx^* \in \bbC^d$ such that
\begin{equation}
    \by = |\bA\bx^*|^2.
    \label{eq:pr_definition}
\end{equation}
There, one assumes that the measurements $\by \in \bbR^n$ and the matrix $\bA \in \bbC^{n \times d}$ are known, while $|\cdot|$ is the element-wise modulus operator (Figure \ref{fig:drawing}(a)). Compared to the linear equation $\by = \bA\bx^*$, the missing phase in \eqref{eq:pr_definition} makes the computational problem more challenging. 

Phase retrieval is a problem that is encountered in many fields. In the context of imaging with electromagnetic waves, it arises whenever one wants to image an object from intensity-only measurements (Figure \ref{fig:drawing}(b)). It has been described in the physics literature since the 1950s, first in crystallography \cite{sayre1952some}, followed by intense studies in astronomy since the '80s \cite{fienup1987phase}. More recently, it has been intimately linked with the next generation of image-reconstruction algorithms which solve this computational problem to push the limits of conventional imaging \cite{miao1999extending, zheng2013wide, yeh2015experimental, wu2018lensless}. Today, phase retrieval encompasses vastly different imaging settings, from X-ray to optical or THz imaging, to observe objects from molecular to astronomical scales. This equation is also encountered outside of computational imaging, for example in computer-generated holography \cite{zhang20173d} or optical computing \cite{chang2018hybrid}, to name a few. Despite the wide range of applications, the underlying physics-based models lead to the same universal phase-retrieval problem. 

The phase-retrieval problem has stimulated the development of a variety of novel imaging modalities. Likewise, a plethora of reconstruction algorithms have been designed \cite{fannjiang2020numerics}, from the first alternating-projection algorithms to convex relaxations \cite{candes2013phaselift} and spectral methods \cite{candes2015phase_wf, luo2019optimal}. Given the rich palette of new reconstruction strategies, it may be unclear for practitioners which one to choose for a particular application. A description of their performance and range of applicability would help in this regard. 

Phase retrieval raises theoretical questions such as the uniqueness of the solution \cite{bandeira2014saving} and the performance guarantees of algorithms \cite{luo2019optimal,mignacco2021stochasticity}. For instance, strong results have been recently obtained for random models where the elements of $\bA$ are drawn as i.i.d.\ random variables. In this case, a clear picture has emerged on regions of solvability \cite{mondelli2019fundamental, maillard2020phase} and the best
algorithm to use. While recent theoretical progress has pushed our understanding to more structured matrices $\bA$ \cite{maillard2020phase}, tackling other models related to real-world settings remains a challenge. 

Phase retrieval is also deeply linked with machine learning. The underlying forward model can be interpreted as a single-layer neural network with quadratic activation function (Figure \ref{fig:drawing}(c)). Thus, the solution to the phase-retrieval problem amounts to the learning of the network weights, which is a typical machine-learning optimization problem. While phase retrieval certainly does not expose one to the full complexity of deep learning, it is already rich enough to provide an entry point to the difficult questions debated in the machine-learning community today. In particular, it exhibits local minima, strongly relies on the quality of the initialization, and behaves differently depending on the amount of training data. As a result, it has sparked a strong interest in the theoretical machine-learning community. In particular, phase retrieval could help us understand better the non-convex optimization procedures typical of real-world machine learning. Ultimately, phase retrieval represents a transdisciplinary topic that impacts fields ranging from imaging applications to fundamental mathematics.

In this tutorial, we present an overview of phase retrieval today, from imaging applications to algorithms and theory. We introduce a unifying framework for the various problems in which a phase-retrieval equation arises: diverse and scattered applications can actually be grouped into four main categories depending on the matrix $\bA$. We also present the latest algorithmic and theoretical advances, in particular, the recent results developed by the theoretical machine-learning community during
the past few years. 
\new{The intuition and principles behind these concepts are described without diving into the technical points. These are described in previous works or dedicated reviews we reference for the interested reader.} We hope that our unified overview will foster interaction between research fields and lead to a better understanding of the phase-retrieval equation. 



\section{Background}

\subsection{Mathematical model}\label{subsec:definitions}

The phase-retrieval equation relates three mathematical objects. In the context of imaging, one wants to reconstruct an object of interest $\bx^*$ from the knowledge of the captured intensities $\by$ and the imaging system modeled by the linear operator $\bA$. Beyond the field of imaging, this equation often arises over the estimation of a physical quantity $\bx^*$ from energy measurements on a sensor, as detectors are often sensitive to the squared modulus of a field. 
It will be useful to rewrite this equation as
\begin{equation}
    \label{eq:pr_definition_scalar_product}
    y_i = |\langle \ba_i, \bx^*\rangle |^2 \quad \text{for } i = 1, \ldots, n,
\end{equation}
where $\bA$ is the vertical concatenation of the rows $\ba_i^{\text{H}}$ 
($\text H$ denoting Hermitian conjugation), 
and $\langle \bx, \by \rangle = \bx^\textrm{H} \by$ is the complex inner product.
Thus, each measurement is obtained from the squared modulus of a scalar product with a sampling vector $\ba_i \in \bbC^d$. 
The phase-retrieval problem can then be restated as follows: How many phaseless scalar products are required to reconstruct the unknown vector $\bx^*$?

In order to make algorithms robust to experimental measurements, the model can be enriched by incorporating a general noise channel $p_\out$ such that $y_i \sim p_\out\left( \cdot \middle| |\ba_i^{\text{H}} \bx^*|^2\right)$, which typically is additive Gaussian or Poissonian. The noise statistics can indeed be leveraged in maximum-likelihood and Bayesian algorithms. 


An important quantity to determine the difficulty of the problem is the \emph{oversampling ratio} $\alpha = n / d$. It compares the number of measurements to the number of parameters we want to estimate. The greater the oversampling ratio $\alpha$, the easier the computational problem becomes. This sampling ratio also plays a role in the simpler problem of linear regression $\by = \bA\bx^*$. There, the solution is not unique---the set of linear equations is underdetermined---for $\alpha < 1$ and can be solved exactly for $\alpha \geq 1$, \new{in the absence of noise and assuming $\bA$ to be full-rank}. Similar considerations apply to phase retrieval, although with a more complicated behavior. In this tutorial, we define and present the latest results regarding injectivity, weak recovery, full recovery, and asymptotic algorithmic performance. These discussions are highlighted in standalone theory boxes. 

Let us emphasize that phase retrieval is, at first sight, a non-convex problem. Indeed, if some $\bx \neq \mathbf{0}$ is in the solution set $\mcS$ of the phase-retrieval equation~\eqref{eq:pr_definition}, then the whole set $\{ e^{i\theta} \bx, \theta \in [0, 2\pi) \}$ is included in $\mcS$, while the middle point $\frac{1}{2} (\bx + (-\bx)) = \mathbf{0}$ is not in  $\mcS$. This global phase ambiguity can be removed for instance by a quotient of the solution space, or by considering instead the estimation of invariant quantities such as $\bx^* (\bx^*)^\text{H}$. 
Notwithstanding, the solution set might still be non-convex. Thus, the general phase-retrieval problem cannot be analyzed directly through the prism of convex optimization.

\subsection{Neural-network representation}\label{subsec:neural_net}

As \eqref{eq:pr_definition} describes a linear operation followed by an element-wise nonlinearity, phase retrieval is equivalent to the training of a single-layer neural network with a quadratic activation function. More precisely, for each $i = 1, \ldots, n$, the input data $\ba_i$ is presented to a network with weights $\bx$, with a target output $\by$, as depicted in Figure \ref{fig:drawing}(c). Training the network for this regression task amounts to the recovery of $\bx^*$ in the phase-retrieval equation. 

This corresponds to the simplest neural network architecture with a nontrivial polynomial nonlinearity. Still, it captures the difficulties of training a shallow neural network with a non-convex loss functional. In particular, convergence is not assured with gradient-descent methods as they may get trapped into local minima or saddle points. 
More generally, single-layer neural networks belong to the class of generalized linear models. This simple yet rich architecture has attracted the attention of theoreticians, which explains the recent profusion of studies around phase retrieval. This parallel enriches the discussion between physics and machine learning, as we shall further illustrate throughout this tutorial. 



\section{Applications}\label{sec:applications}

\subsection{Unifying Framework for Phase-Retrieval Applications}\label{subsec:unifying_framework}

The phase-retrieval problem is central to a large variety of applications in science and technology. Although many of these approaches were developed independently over the years, they share the same phase-retrieval equation~\eqref{eq:pr_definition}.
Diversity comes notably from the set of computations performed by the matrix $\bA$ that depends on the physical model of the imaging system. This domain knowledge sometimes makes it difficult for non-experts to navigate the landscape of phase-retrieval applications.

To offer a unifying perspective, we are proposing to classify phase retrieval into four major groups: Fourier, coded-illumination, coded-detection, and random models (Figure \ref{fig:drawing}(d)). Targeted toward imaging, this classification has been designed to facilitate interaction between fields. It may be used by practitioners to draw links between applications and benefit from the expertise of other fields. For theoreticians, it may give directions to follow when defining new theoretical models to study.

\subsection{Fourier Phase Retrieval}\label{subsec:fourier_pr}

Fourier phase retrieval has been historically the first occurrence of this nonlinear equation. Wave propagation over a long distance, referred to as far-field propagation, is modeled by a Fourier transform \cite{goodman2005introduction}, as in $\bA = \bF$, where $\bF$ is a 2D Fourier-transform matrix. The first phase-retrieval problems appeared in modalities where optical components are difficult to manufacture, most notably for X-rays or electron imaging \cite{miao1999extending}. In these cases, the imaging system is often minimalist and only involves far-field propagation. From the determination of molecular structures using X-rays to aberration correction in astronomy \cite{fienup1987phase}, Fourier phase retrieval has been associated with major scientific discoveries in the past century. 

Since the Fourier transform is a unitary operation (i.e.\ $\bF^{\text{H}} = \bF^{-1}$), we necessarily have that $n = d$, leading to the oversampling ratio $\alpha = 1$. This problem is fundamentally ill-posed as we only know amplitudes with no phase information about the Fourier transform of $\bx^*$. Additional information on $\bx^*$ is required to lift this degeneracy. The famous work of Gerchberg and Saxton \cite{gerchberg1972practical} uses another intensity measurement of $\bx^*$ directly so that, in \eqref{eq:pr_definition}, $\bA$ becomes a vertical concatenation of the identity and Fourier operators. Another common assumption exploits a support and non-negativity constraint on $\bx^*$ \cite{fienup1987phase}, which is often valid in astronomy. \new{Other more involved assumptions may be invoked, and they will be described in Section \ref{sec:regularization} dedicated on regularization. New applications of Fourier phase retrieval have recently appeared, such as speckle correlation imaging for non-line-of-sight imaging or complex media imaging \cite{katz2014non,metzler2020deep}.} Ultimately, while Fourier phase retrieval has been the first physical model of phase retrieval, it remains a challenging computational problem. 



\subsection{Coded-illumination Phase Retrieval}\label{subsec:coded_illumination}

More advanced imaging techniques were developed to overcome the overly restrictive assumptions of Fourier phase retrieval. A solution to lift the phase ambiguity consists in shining different illuminations onto the object of interest in order to capture multiple images after far-field propagation. Each image corresponds to a slightly different view of the same object. This provides the algorithm with complementary and partially redundant information and improves its convergence, at the expense of a more involved experimental setup and longer measurement time. This is the general principle underlying \textit{coded-illumination phase retrieval}. Here, the linear operator $\bA$ is a vertical concatenation of operators $\bA_l$:
\begin{equation}
    \bA_l = \bF \bD_l \text{ for }l = 1, \ldots, L,
\end{equation}
where $\bD_l$ is a diagonal matrix that corresponds to the coded illumination, $n = L d$, and the oversampling ratio $\alpha$ is equal to the number $L$ of images.
A variety of imaging applications based on coded illuminations are listed below. 
\begin{itemize}[leftmargin=20pt]
    \item Coded diffraction imaging relies on a modulation near the object plane \cite{candes2015phase}. This modality allows ample flexibility on the choice and implementation of the modulation scheme, which explains its application in many different imaging settings. Gratings, masks, as well as tilted wavefronts all represent solutions for structured illumination. 
    \item Ptychography is arguably the most successful implementation of coded-illumination phase retrieval \cite{rodenburg2019ptychography}. It records the diffraction pattern of the object while it is scanned by an illumination probe with finite support. Oversampling is linked with the overlap between two consecutive illuminated areas---the higher the overlap, the easier it is to solve the phase-retrieval problem. It has been successfully used with a broad range of electromagnetic waves, from X-rays to the THz regime, as well as for electron microscopy. 
    \item Fourier ptychography is a variant of ptychography in Fourier space where a sequence of tilted plane waves are sent on the sample \cite{zheng2013wide, yeh2015experimental}. In this case, the object $\bx^*$ is defined as the Fourier transform of the object transmission function, and the diagonal matrix $\bD_i$ corresponds to a finite-support probe that is translated in Fourier space. This technique is used in optical microscopy to achieve high-resolution on a large field-of-view.
\end{itemize}
Imaging scientists have developed specialized algorithms for each of these applications. For example, the Ptychographic Iterative Engine \cite{rodenburg2019ptychography} is a popular projection algorithm, while gradient-based methods are also routinely used. However, the structure present in these sensing matrices prevents a thorough establishment of theoretical results for now.



\subsection{Coded-detection Phase Retrieval}\label{subsec:coded_detection}

Acting on the illumination in the proximity of the object may not always be feasible, for instance in astronomical observations.
Nevertheless, multiple images can be obtained upon modulation on the detection side, leading to \textit{coded-detection phase retrieval}.
This can be done by taking several shots at different defocused positions---a technique derived again from astronomy and called phase diversity \cite{paxman1992joint}, but applied to other imaging fields as well, such as lensless imaging \cite{wu2018lensless}. 

There are many ways to describe defocus via free-space wave propagation, and all rely on various degrees of approximation of Maxwell's equations. Using the angular-spectrum method, the operator $\bA$ is now the vertical concatenation of
\begin{align}
    \bA_l = \bF \bD_l \bF^{\text{H}} \text{ for }i = 1, \ldots, L, \\
    \textrm{ with } \bD_l = \operatorname{Diag} \left(e^{i z_l \sqrt{1-|\mathbf{u}|^2}}\right), \nonumber
\end{align}
where $\bu$ denotes normalized coordinates in Fourier space and the unitless curvature $z_l$ is proportional to the defocus distance. Interestingly, this class of phase retrieval also appears outside of imaging for computer-generated holography \cite{zhang20173d}. There, one wants to produce a 3D hologram---a volumetric intensity pattern. This target defines a set of equations at multiple planes, such that a phase mask is optimized to satisfy these constraints and applied with a wavefront-shaping device. 

More general coded-detection strategies may be designed to simplify the phase recovery by using the diversity of the diagonal matrices $\bD_l$ in Fourier space. For example, one can add higher-order optical aberrations beyond defocus or a physical aperture. Once again, a thorough theoretical understanding of this phase-retrieval class is still lacking. It would especially benefit from more results regarding solvability and algorithm choice.

\subsection{Random Phase Retrieval}\label{subsec:random_models}

In random phase retrieval, $\bA$ is a random matrix. A common setting is the Gaussian model, in which each element of $\bA$ is drawn from an i.i.d.\ complex Gaussian distribution. Such an operator $\bA$ may be seen as a generic linear operation: each sampling vector is independent from the others and adds a little bit of information on the unknown $\bx^*$. Contrary to expectations, random phase retrieval arises in practical applications too, such as imaging in complex media \cite{metzler2017coherent}. There, the random linear operator comes from multiple light scattering or random input patterns synthesized by a programmable device. 

Distinct from the previous models, randomness greatly facilitates theoretical studies and allows one to derive asymptotic results on recovery guarantees and to characterize the typical behavior of algorithms. These advances are recent and will be described in more details below. The bottomline is that phase retrieval in the random setting is quite well-understood with active research still ongoing.

Some imaging applications exhibit pseudo-random models, such as coded diffraction imaging with random masks or ptychography with a random probe \cite{rodenburg2019ptychography}. They emulate random models, but residual correlations are present between rows of $\bA$. This variable amount of randomness makes it challenging to derive precise theoretical guarantees on performance but, to some extent, motivates the use of the latest algorithms designed for the fully random setting. 

These observations motivate further work to bridge the gap between experiments and theory. Indeed, to illustrate current efforts, a few theoretical results have recently been generalized beyond the i.i.d.\ random setting \cite{maillard2020phase}. We can distinguish between two classes of random models, for which strong theoretical results have been reported
\begin{enumerate}[label=\textbf{R.\arabic*},ref=R.\arabic*]
\item \label{model:gaussian_iid} \emph{(i.i.d.\ random)} The matrix components of $\bA$ satisfy
$[\bA]_{i,k} \overset{\mathrm{i.i.d.}}{\sim} P$, with $P$ a complex (centered) probability distribution with finite moments of all order (for simplicity).
\item \label{model:rot_inv} \emph{(right unitarily-invariant)}
$\bA$ is a random matrix with probability distribution $P(\bA)$, and $P(\bA) = P(\bA \bU)$ for every unitary matrix $\bU$. Equivalently, the right eigenvectors of $\bA$ are \emph{completely delocalized}.
%
\end{enumerate} 
As a final remark, the observed robustness of random phase retrieval may inspire new ideas to
introduce randomness in experimental settings.

\floatstyle{boxed} 
\restylefloat{figure}
\begin{figure*}
    \begin{minipage}{1.\textwidth}
        \textbf{Injectivity: When is the solution unique?}
        
        In order to build a mathematically sound theory of phase retrieval, 
        it is important to understand if the solution to the (noiseless) phase-retrieval problem is \emph{unique}. Stated differently, for a given measurement matrix $\bA \in \bbC^{n \times d}$, is the map $\mcA_\bA: \bx \in \bbC^d \to |\bA \bx|^2 \in \bbR^n$ injective\footnote{While we do not mention it explicitly, the reader should remember that injectivity is always meant up to a global phase.}?
        Since 2014, the following picture has gradually emerged on a complete characterization of injectivity.
        \begin{conjecture}\label{conj:4dm4}
            For any $d \geq 2$, and any full-rank $\bA \in \bbC^{n \times d}$
            the following holds
            \begin{itemize}
                \item[$(i)$] If $n < 4d-4$, then the map $\mcA_\bA$ is non-injective. 
                \item[$(ii)$] If $n \geq 4d-4$, then the map $\mcA_\bA$ is injective for a ``generic'' $\bA$\footnote{We refer to the description of the conjecture in \cite{bandeira2014saving} for a more detailed definition of the notion of ``generic'' $\bA$.}.
            \end{itemize}
        \end{conjecture}
        First described as the ``$(4d-4)$ conjecture'' \cite{bandeira2014saving}, point~$(ii)$ was later proven in \cite{conca2015algebraic}. Meanwhile, point $(i)$ has been proven for small dimensions $d \leq 3$ in \cite{bandeira2014saving}, disproved for $d = 4$ and refined for large dimensions in a probabilistic sense\footnote{Let $n = 4d - 5$.
        Let $\mathrm{im}(\bA)$ be drawn at random from the set 
        of $d$-dimensional subspaces of $\bbC^n$,
        and denote $p_d$ the probability that $\mcA_\bA$ is injective
        . Then:
        $(i)$ for all $d \geq 2$, one has $p_d < 1$, and
        $(ii)$ $\lim_{d \to \infty} p_d = 0$.} \cite{vinzant2015small}.
        
        In particular, for large $d \gg 1$, the threshold $n/d \simeq 4$ should indeed distinguish between ``typically injective'' and ``typically non-injective'' behaviors.
        In practice, in order to efficiently recover in the not-too-oversampled regime $n/d \lesssim 4$,
        this suggests that one should leverage \emph{a priori} information on the object (e.g.\ regularization methods).
        \label{box:injectivity}
    \end{minipage}
\end{figure*}
\floatstyle{plain} 
\restylefloat{figure}

\section{Reconstruction Algorithms}\label{sec:algorithms}

A large number of reconstruction algorithms have been proposed to tackle the diversity of phase-retrieval models. We present them here together with some intuitive explanation and indications on when to use them in practice. The purpose of this overview is to illustrate the inspiring creativity that researchers deployed while designing phase retrieval algorithms. Dedicated surveys cover more technical descriptions of the recent developments \cite{fannjiang2020numerics, shechtman2015phase}.

\subsection{Projection Algorithms} \label{subsec:projection}
    
Projection algorithms were the first strategy proposed to solve the phase-retrieval problem. They provide a direct and intuitive way to exploit prior information or several intensity measurements. Thanks to their ease of implementation and flexibility, projection algorithms are still widely used today. 

More precisely, each intensity image or prior knowledge corresponds to a constraint labelled from $1$ to $N$. The earliest implementation is the Gerchberg-Saxton algorithm \cite{gerchberg1972practical}, which uses the intensity at two measurement planes as constraints in the object and Fourier planes. The error-reduction algorithm resorts to a modulus constraint in the Fourier plane and a support constraint in the object plane \cite{fienup1987phase}. Other techniques such as ptychography \cite{maiden2009improved, rodenburg2019ptychography}, Fourier ptychography \cite{zheng2013wide, yeh2015experimental}, or phase diversity perform projections onto more than two sets to exploit even more information for to recover the missing phase. 

Each constraint is satisfied by a set of vectors $\bx \in \bbC^d$, and one is aiming at finding the intersection of such sets. One thus projects on all the sets sequentially, to hopefully converge to a suitable solution. The initial estimate is $\bxhat^{(0)}$ and the iterations proceed according to
\begin{equation}
    \label{eq:projection_algorithm_update}
    \bxhat^{(k+1)} = \Pi_N\circ\ldots\circ\Pi_1(\bxhat^{(k)}),
\end{equation}
where $\Pi_l$ denotes the projection operators on the solution set of constraint $l = 1, \ldots, N$.

For example, we can write the projector of a modulus constraint for an intensity image $\by_l = |\bA_l \bx^*|^2$ that corresponds to an invertible linear operator $\bA_l$ as
\begin{equation}
    \label{eq:projector_modulus}
    \Pi_l(\bu) = \bA_l^{-1} \Diag \left(\frac{\sqrt{\by_l}}{|\bA_l \bu|}\right) \bA_l \bu.
\end{equation}
This operator effectively replaces the square modulus of $\bA_l \bu$ by the one obtained from the measurements $\by_l$. Similarly, the projector of a support constraint $\gamma$ sets all the values outside the support to zero: $\Pi(\bu) = \Diag(\indi_\gamma) \bu$, where $\indi_\gamma$ denotes the indicator function of $\gamma$. 

While this class of algorithms is widely used in practice, it unfortunately does not necessarily converge to the set of solutions. More advanced algorithms have been suggested for better convergence, such as the hybrid input-output algorithm \cite{fienup1987phase} that uses relaxed projections for Fourier measurements with support constraint. Another strategy replaces the sequential projection with an average to avoid being stuck in loops. When applied to ptychographic algorithms, these concepts gave rise to the difference map and relaxed averaged alternating reflections algorithms \cite{rodenburg2019ptychography}. 

\subsection{Gradient-based Optimization}\label{subsec:gradient_optimization}

Gradient-based iterative methods are the methods of choice to optimize nonlinear objective functions in machine learning. This class of algorithms is very flexible, with many variations and acceleration strategies. They have been used to solve the phase-retrieval problem \cite{guizar2008phase, candes2015phase_wf}, with applications ranging from ptychography and coded-diffraction imaging to defocus-based implementations.  

One wants to find an estimate $\bxhat$ that minimizes a loss function $\mcL(\bx, \by)$, for instance the square loss $\mcL_2(\bx, \by) = \left\|\by - |\bA \bx|^2\right\|^2$. 
At each iteration $k$, a step of size $\eta_k$ toward the negative gradient of $\mcL$ is taken to refine the solution, leading to the process
\begin{equation}
    \label{eq:gradient_descent}
    \bxhat^{(k+1)} = \bxhat^{(k)} - \eta_k  \nabla_{\bxhat} \mcL(\bxhat^{(k)}, \by).
\end{equation}
The choice of the loss function is of crucial importance since it determines the performance of the algorithm. An educated guess of a suited loss function can be made if the noise statistics are known. Typical noise models are Poissonian (for shot-noise limited measurements) or additive Gaussian \cite{yeh2015experimental}. In this case, the loss function corresponds to the negative log-likelihood function $-\log(p_\out\left(\by \middle| |\bA \bx^*|^2\right))$. Typically, the loss function $\mcL_2$ corresponds to the negative log-likelihood of the-additive-Gaussian noise case.

As the optimization is non-convex, gradient descent may get stuck in local minima or saddle points. Many acceleration strategies are at our disposal to circumvent these difficulties, from conjugate gradient and Nesterov acceleration to second-order methods. Another classical technique in machine learning to escape local minima is stochastic gradient descent, in which the gradient of the loss function at each iteration is computed using a partial dataset only. Intensity images are often used one at a time, akin to sequential projection algorithms, although this introduces noise in the gradient-descent updates.

Gradient descent represents a preferred tool to solve the phase-retrieval equation. Nevertheless, the theoretical bases of gradient-based algorithms still represent an active research topic. For example, the choice of the initial estimate greatly impacts the performance: in Section~\ref{subsec:spectral} we shall describe
algorithms that yield informed initializations.

\subsection{Convex Relaxations}\label{subsec:convex_relaxation}

In contrast to gradient descent, which is a universal optimization technique, convex relaxations exploit the quadratic form of the phase-retrieval equation.

Convex relaxations recast the phase-retrieval problem using auxiliary variables. We seek a slightly different or \textit{relaxed} formulation of the problem, which makes it convex. Convex optimization then provides a large variety of algorithms, with stronger convergence guarantees.
For instance, PhaseLift \cite{candes2013phaselift} replaces the unknown vector $\bx^* \in \bbC^d$ by the matrix $\bX^* = \bx^*(\bx^{*})^{\text{H}} \in \bbC^{d \times d}$. The quadratic equation~\eqref{eq:pr_definition_scalar_product} then becomes linear in the components of $\bX^*$ since one now has that
\begin{equation}
    y_i = \ba_i^{\text{H}} \bX^* \ba_i \quad \text{ for } i = 1, \ldots, n.
\end{equation}
In its original form, the phase-retrieval equation would require one to seek a rank-one solution. As this rank constraint is itself non-convex, PhaseLift looks for a minimal-trace solution instead, leading to the convex program
\begin{align}
    \text{minimize} \, \, \operatorname{Tr}(\bX), 
    \quad 
    \textrm{subject to } 
    \bX \geq 0 
    \\ \nonumber
    \text{and } 
    \quad
    y_i = \ba_i^{\text{H}} \bX \ba_i 
    \quad 
    \text{ for } i = 1, \ldots, n.
\end{align}
Written in this form, the problem falls into the category of semidefinite programming, where one can readily use an off-the-shelf solver. The output is then given by the leading eigenvector of the final estimate $\bX$. Other convex relaxations have been developed \cite{waldspurger2015phase, goldstein2018phasemax}, \new{a detailed list being reviewed in} \cite{fannjiang2020numerics}. 

The issue associated to semidefinite programming lies in its computational complexity, as one needs to build a matrix in $\bbC^{d \times d}$. To illustrate this issue, consider a $100 \times 100$ image, which would correspond to $d = 10^4$. Standard convex-relaxation techniques would then require one to perform optimization on a $d \times d = 10^8$-dimensional space. For this reason, semidefinite-programming strategies have only been used as a proof of concept in real imaging settings, although promising sketching methods to keep the computation tractable have been proposed recently \cite{yurtsever2017sketchy}. 



\subsection{Spectral Methods}\label{subsec:spectral}

To enhance the performance of the possibly expensive methods described in the previous sections, practitioners often require an ``informed'' initialization $\bxhat^{(0)}$, preferably one that is close to optimum. Recently proposed, the privileged class of algorithms to obtain such initializations at a low computational cost are \emph{spectral methods} \cite{candes2015phase_wf}. 
These methods build an estimate $\bxhat$ as the principal eigenvector of a weighted covariance matrix, defined from the intensities $\{y_i\}$ and the sampling vectors $\{\ba_i\}$ as
\begin{equation}\label{eq:def_spectral_method}
    \bM(\mcT) = \frac{1}{n} \sum_{i=1}^n \mcT(y_i) \ba_i \ba_i^{\text{H}}.
\end{equation}
Intuitively, since $y_i = |\ba_i^{\text{H}} \bx^*|^2$, this scalar measurement is minimal when $\ba_i$ and $\bx^*$ are orthogonal, while it increases when the two vectors are correlated. For any increasing \emph{preprocessing function} $\mcT$, the matrix in \eqref{eq:def_spectral_method} thus gives ``more weight'' to those sampling vectors $\ba_i$ that are more aligned with $\bx^*$.

The performance of a given spectral method heavily depends on the choice of the preprocessing function $\mcT$. Various choices have been considered in the literature, for example the trimming scheme $\mcT(y) =  y \indi\{|y| \leq t\}$ (with possibly $t = \infty$, meaning in this case that $\mcT$ is the identity). 
Interestingly, an optimal preprocessing method has been derived for the random models \ref{model:gaussian_iid} and \ref{model:rot_inv} and any type of noise \cite{luo2019optimal, maillard2021construction}. In particular, in a noiseless setting it reads\footnote{In order to simplify our analysis, we assume (without loss of generality) the scaling $\lVert \bx^* \rVert^2 = 1$ and $(1/n) \Tr[\bA \bA^\text{H} / d] = 1$.
} $\mcT^\star(y) = 1 - y^{-1}$. 

In practice, the simplicity of spectral methods ensures that their computational cost remains very low. The leading eigenvector can basically be recovered with power iterations or its accelerated variants\footnote{For instance, \cite{luo2019optimal,maillard2021construction} show that the largest eigenvalue of $\bM(\mcT^\star)$ concentrates on $\lambda_\mathrm{max} = 1$ when recovery is possible, so that one can use very efficient inverse iterations, i.e.\ power iterations on $(\bM(\mcT^\star) - \Id)^{-1}$.}. Importantly, the convergence toward the leading eigenvector is ensured, a guarantee that is missing in many other nonlinear optimization strategies. Unlike convex relaxation methods, spectral methods do not involve a lifting process either, thus simplifying their adoption in imaging settings such as ptychography \cite{valzania2021accelerating}. The solution estimated via spectral methods is then typically refined with an iterative optimization algorithm such as gradient descent.

\begin{figure*}
    \centering
    \includegraphics[width=\linewidth]{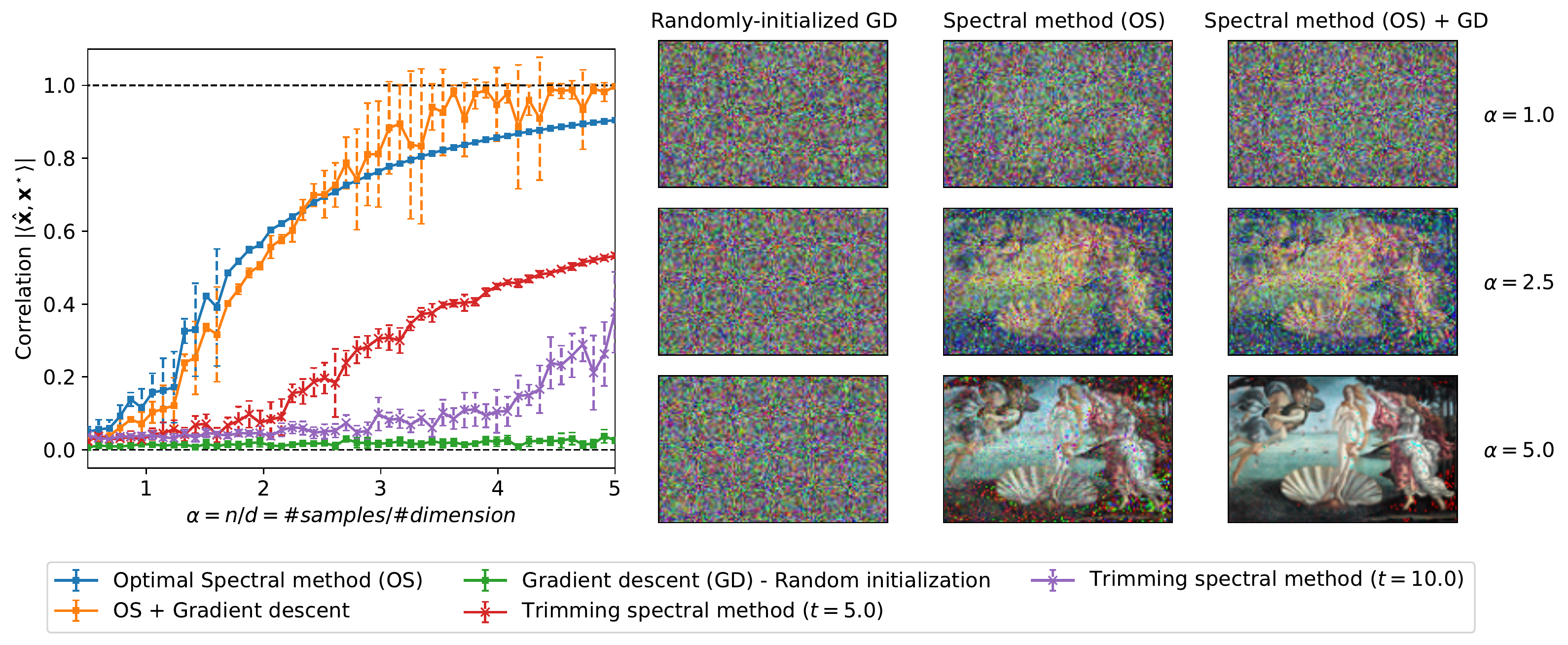}
    \caption{Correlation $|\langle \bxhat, \bx^* \rangle|$ achieved by spectral methods in noiseless phase retrieval with a complex Gaussian matrix, for the recovery of an image.
    The Optimal Spectral method (OS) is compared with the trimming method with different values of $t$.
    Gradient descent is initialized with the OS, showing zero error already at $\alpha \gtrsim 4$. Finally, a comparison with randomly initialized gradient descent is shown, which does not achieve a positive correlation at any $\alpha \leq 5$. 
    In the last two columns, the optimal spectral method was chosen.
    }
    \label{fig:spectral_methods}
\end{figure*}

In Figure~\ref{fig:spectral_methods} we illustrate the performance of some implementations of spectral methods, in the noiseless phase-retrieval problem with $\bA$ being a complex Gaussian matrix. The optimal spectral method $\bM(\mcT^\star)$ yields almost perfect correlation with the true solution $\bx^*$, even without subsequent refinement by gradient descent. Results are also compared with bare gradient-descent reconstructions. They show how all these theoretical findings transfer very well to the recovery of a real image.

\floatstyle{boxed} 
\restylefloat{figure}
\begin{figure*}
    \begin{minipage}{1.\textwidth}
        \textbf{Weak Recovery Threshold: When do reconstructions start to be possible?}
        
        Complementary to injectivity studies, many recent works have focused on the question of \emph{weak recovery}, in other words
        when does it become possible to \emph{nontrivially} estimate the signal $\bx^*$? In mathematical terms, this means that there exists a procedure 
        that is polynomial-time in $d$ and
        that returns an estimator $\hat{\bx}$ satisfying
        \begin{equation}\label{eq:weak_recovery}
            \lim_{d \to \infty} |\langle\hat{\bx} , \bx^* \rangle| > 0.
        \end{equation}
        Note that an $\bxhat$ uniformly sampled on the sphere (i.e.\ completely random) satisfies with high probability $|\langle \bxhat, \bx^* \rangle| = \mathcal{O}(1/\sqrt{d}) \to 0$ for $d \to \infty$. 
        
        A series of precise results on the weak recovery of phase retrieval under a random design assumption were 
        obtained in the recent years. 
        In particular, under \ref{model:gaussian_iid} or \ref{model:rot_inv}, 
        \cite{mondelli2019fundamental,maillard2020phase} showed that if $n, d \to \infty$ with $\alpha = n/d$, there exists a sharp threshold $\alpha_\WR$ distinguishing between impossible ($\alpha < \alpha_\WR$) and possible
        ($\alpha > \alpha_\WR$) weak recovery.
        %
        %
        For example, in noiseless complex phase retrieval with a Gaussian measurement matrix, this yields $\alpha_\WR = 1$, as seen in Figures~\ref{fig:spectral_methods} and \ref{fig:amp}.\footnote{\cite{maillard2020phase} provides an explicit equation solved by $\alpha_\WR$, solely as a function of the \emph{singular-value spectrum} of the measurement matrix $\bA$. For example, in complex noiseless phase retrieval, $\alpha_\WR$ is the only solution to $\alpha = 2 (\EE (\lambda))^2 / \EE (\lambda)^2$,
        in which $\EE(\cdot)$ is taken with respect to the eigenvalues $\{\lambda\}$ of $\bA \bA^\text{H} / d$. Note that the right-hand side depends on $\alpha$ in general, making this an implicit equation.}
        Note that the ``optimal'' spectral method we described in Section~\ref{subsec:spectral}
        has been shown to reach the asymptotic weak-recovery threshold, i.e.\ it satisfies Eq.~\eqref{eq:weak_recovery} 
        for any $\alpha > \alpha_\WR$ \cite{mondelli2019fundamental,maillard2021construction}.
    \label{box:weak_recovery}
    \end{minipage}
\end{figure*}
\floatstyle{plain} 
\restylefloat{figure}

\subsection{Bayesian Algorithms}\label{subsec:bayesian_algos}

Bayesian estimators are a widely used and powerful ensemble of algorithms that leverage the properties of the \emph{posterior} distribution, which is given by Bayes' law:
\begin{equation}\label{eq:posterior}
    \bbP(\bx | \by, \bA) = \frac{p_\out(\by | |\bA \bx|^2) p_0(\bx)}{\bbP(\by | \bA)}.
\end{equation}
Here, $\by$ are the observations, $p_\out$ is the \emph{likelihood}, and $p_0$ a \emph{prior} distribution, typically Gaussian if one does not want to enforce any particular structure on the reconstructed image. 
Bayesian algorithms are not based on the minimization of a loss function, but rather on the computation of estimates defined from the posterior distribution. Thus, they can potentially exploit more efficiently the available information in the measurements. 
A typical Bayesian estimator is the Minimal Mean-Squared Error estimator $\bxhat_\MMSE = \argmin_\bx \EE_{\bx'}[\lVert\bx' - \bx\rVert^2 | \by, \bA] = \EE [\bx | \by, \bA]$.
However, the computation of such estimators---defined on the posterior distribution---is a hard problem in general, since sampling from the posterior is often an exponential-time procedure with classical Monte Carlo Markov Chain (MCMC) approaches.

In this regard, an important class of iterative Bayesian algorithms is approximate message passing (AMP) \cite{schniter2016vector}, which is based on high-dimensional asymptotics of the \emph{belief-propagation} algorithm \cite{mezard2009information}. 
The power of this class of algorithms arises from several insights. First, AMP is known to be optimal among a large class of polynomial-time algorithms for many random models \cite{celentano2020estimation} and is conjectured to be so for many more. Moreover, it is well understood, in the sense that one can analytically track its performance in the high-dimensional limit. However, AMP is computationally heavy compared to simpler algorithms, such spectral methods or gradient-based optimization procedures. 
More generally, the main limitation of Bayesian algorithms is that they assume the knowledge of the generative model (i.e.\ of the model of the noise present in the observations, and of the distribution of the random sensing matrix).
While algorithmic procedures such as AMP can be used off-the-shelf beyond these assumptions (e.g.\ for a non-random $\bA$), their performance and convergence properties are no longer guaranteed\footnote{While AMP achieves optimal polynomial-time performance in phase retrieval with random unitary matrices $\bA$, it performs very poorly when $\bA$ is a vanilla Fourier matrix \cite{maillard2020phase}.}. 

\begin{figure*}
    \centering
    \includegraphics[width=\linewidth]{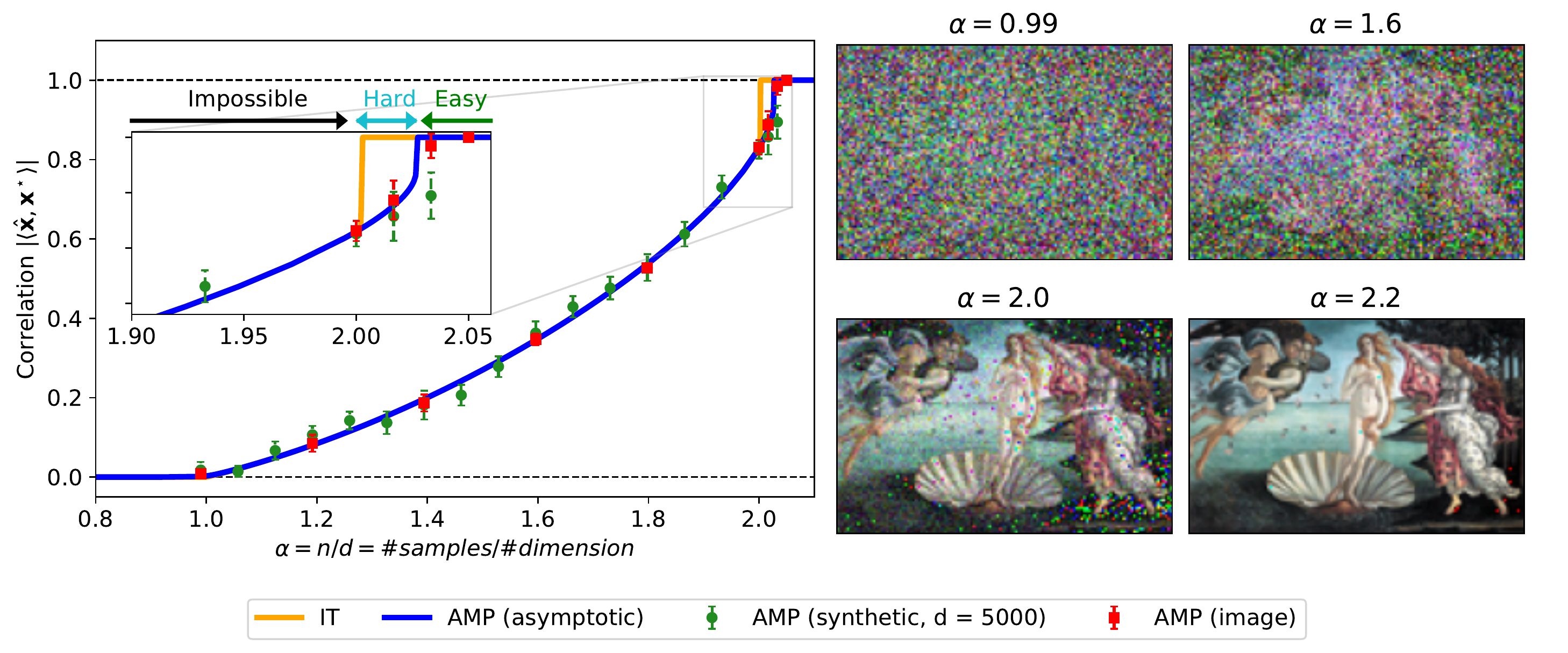}
    \caption{Correlation $| \langle \bxhat, \bx^* \rangle|$ achieved by AMP in noiseless phase retrieval with a complex Gaussian matrix.
    We compare the theoretical prediction, known as ``state evolution'' (blue curve), with actual runs of the algorithm.
    The results are very consistent whether the signal $\bx^*$ is synthetic (i.e.\ random) or an actual image, achieving perfect recovery for $\alpha \gtrsim 2.03$.
    The inset of the left column compares the performance of AMP with the information-theoretic (IT) performance, which is the optimal error that any algorithm can reach, no matter its running time. 
    We uncover a \emph{computational gap} (or ``hard phase'') for $2 < \alpha \lesssim 2.03$, for which AMP is not able to reach the IT performance \cite{maillard2020phase}.
    We use an efficient modular implementation of AMP in a wide class of inference procedures (including phase retrieval), publicly available in \cite{baker2020tramp}.
    }
    \label{fig:amp}
\end{figure*}

We show in Figure~\ref{fig:amp} the performance of AMP in noiseless phase retrieval with a complex Gaussian matrix. We obtain a very good agreement with the theory (box on asymptotic guarantees) and, as for the spectral methods \cite{maillard2021construction}), a very consistent performance for the recovery of real images.

\floatstyle{boxed} 
\restylefloat{figure}
\begin{figure*}
    \begin{minipage}{1. \textwidth}
    \textbf{Perfect recovery: When can we reconstruct ``almost exactly?''}
    
    Crucially, injectivity studies inherently consider ``worst-case'' scenarios on the choice of the signal $\bx^*$ to recover and the measurement matrix $\bA$, and do not leverage the high dimensionality of the problem.
    In order to go beyond this framework, the \emph{perfect recovery} (or full recovery) phenomenon also attracted interest in the past years. 
    This question is fundamentally high-dimensional and can be formulated as follows: \emph{for $d \gg 1$, when is $\alpha = n/d$ large enough 
    so that, for any typical} $\bx^*$ \emph{and} $\bA$\emph{, there exists an estimator} $\bxhat$ \emph{of} $\bx^*$ \emph{(with} $\lVert\bxhat\rVert = 1$\emph{) that satisfies} $|\langle\hat{\bx} ,\bx^*\rangle| \simeq 1$?
    
    Leveraging prior knowledge on the structure of the signal $\bx^*$ (i.e.\ giving a sense to the typicality), the question above has been answered in 
    the very large class of random models \ref{model:gaussian_iid}-\ref{model:rot_inv} \cite{maillard2020phase}.
    In particular, perfect recovery can be achieved much before the typical injectivity threshold $\alpha = 4$, e.g.\ at $\alpha \simeq 2$ for Gaussian matrices $\bA$ and 
    a uniformly-sampled $\bx^*$, cf.\ Fig.~\ref{fig:amp}.
    
    Perfect-recovery studies turn out to be more relevant to practical phase-retrieval setups than injectivity analyses.
    In these setups, the measurement matrix and signal are not worst-case but rather constrained by the properties of the physical problem (e.g.\ a natural image). Then, one is often willing to accept vanishingly small errors.
    \end{minipage}
    \label{box:perfect_recovery}
\end{figure*}
\floatstyle{plain} 
\restylefloat{figure}


\floatstyle{boxed} 
\restylefloat{figure}
\begin{figure*}
\begin{minipage}{1. \textwidth}
\textbf{Asymptotic guarantees: Can we characterize the performance of reconstruction algorithms?}

A considerable effort has been devoted to characterize the performance of the diverse algorithms presented in Section~\ref{sec:algorithms}, with a particular emphasis on the theoretical analysis of the error for large $n, d$,
often under random-design assumptions (cf.\ Section~\ref{subsec:random_models}).

For instance, the Bayesian algorithms of Section~\ref{subsec:bayesian_algos} and, in particular, AMP (which conjecturally reaches the optimal polynomial-time performance) can be analytically studied in the high-dimensional limit with a set of tools known as \emph{State Evolution} \cite{maillard2020phase, schniter2016vector}. On one hand, these allow us to establish the theoretical solid lines in Figure~\ref{fig:amp}, which agree very well with finite-size simulations of AMP.
On the other hand, the spectral methods described in Section~\ref{subsec:spectral} can be tackled more directly using tools of random-matrix theory and the authors of \cite{luo2019optimal} derive analytical formulas for the asymptotic performance achieved by a wide class of spectral methods.
Finally, gradient-based algorithms (Section \ref{subsec:gradient_optimization}) can also be analyzed in high dimension. For instance, \cite{sarao2020optimization} strikingly explained how over-parameterization greatly enhances their performance in phase retrieval. Another avenue to understand the trajectories of 
(stochastic) gradient descent in general settings is \emph{Dynamical Mean Field Theory} \cite{mignacco2021stochasticity,celentano2021high}.

We summarize in Table~\ref{table:asymptotic_guarantees} these performance analyses in the specific case of noiseless complex-valued phase retrieval with the Gaussian/i.i.d.\ model~\ref{model:gaussian_iid}. 
Note that many of these results have been extended to more general noise and measurement models.

Despite missing pieces and several simplifying assumptions, theoreticians have reached a very precise understanding of the performance of phase-retrieval algorithms.
Nevertheless, almost all these studies rely on high-dimensional random models which, even though they can be quite refined, fail to describe 
many practical setups such as coded diffractive imaging with structured Fourier matrices
or low-dimensional configurations.
In particular, the theory of Bayesian methods and of spectral methods has been empirically shown to fail when the eigenspaces of $\bA$ are very structured (e.g.\ with vanilla Fourier matrices).
In this sense, theoretical asymptotic guarantees for structured matrices are still an open problem. 
\label{box:asymptotic}

\end{minipage}
\end{figure*}

\new{
\section{Regularization in the Deep-Learning Era} \label{sec:regularization}

\subsection{Conventional Regularization}
\label{subsec:conventional regularization}

When faced with a difficult phase-retrieval problem, an appropriate regularization can be an efficient solution to enhance recovery of the desired images. It introduces prior information on the class of images to recover, which will help the reconstruction algorithm. This concept has long been associated with phase retrieval, since the first projection algorithms were based on support and positivity constraints.

Most regularization approaches can be formulated as the optimization problem
\begin{equation}\label{eq:minreg}
   \argmin_{\bx\in\bbC^d} \; \{\mcL(\bx, \by) + \mcR(\bx)\}.
\end{equation}
The first term enforces consistency with the intensity measurements $\by$ (data-fidelity term). It corresponds to the metric in the gradient-based reconstruction algorithms discussed in Section \ref{subsec:gradient_optimization}. The second term penalizes unrealistic estimates to favor suitable properties on $\bx$. In the Bayesian setting of \eqref{eq:posterior} with maximum-likelihood estimation, the two terms correspond to $p_\out$ and $p_0$, respectively. In Section \ref{subsec:bayesian_algos}, the prior was set to a generic Gaussian distribution, but it can be modified to introduce other prior information on the recovered image. 

Indicator functions may be used to enforce nonnegativity or support constraints. 
Another frequently studied regularization term is $\|\bx \|_1$, the $L_1$ norm of $\bx$. It promotes sparsity in the final reconstruction and bears deep links with the field of compressed sensing. Sparsity in a transform domain can often be interesting for images, such as sparsity of $\bL\bx$ with $\bL$ the discrete gradient operator for total-variation regularization. 
To tackle these potentially non-differentiable composite-optimization problems, proximal algorithms alternatively minimize the loss $\mathcal{L}$ and the regularization $\mathcal{R}$ using gradients or projection operators. A large variety of algorithms have been designed for this task. We refer the reader to the tutorial \cite{shechtman2015phase} on conventional regularization and sparsity for more details. 

Machine learning algorithms and, in particular, convolutional neural networks (CNNs), have recently emerged as powerful new tools to analyze natural images. 
Remarkably, they achieve state-of-the-art performance on many image benchmarks such as classification, object detection, denoising, or image generation. Not surprisingly, CNNs have also been deployed for image reconstruction as they can provide a very strong prior for natural images. This prior consistently outperforms classical regularization on natural images. There are several ways to introduce it in our phase-retrieval algorithms. 

\subsection{Plugging-in a Denoiser}
\label{subsec:PnP}

CNNs are remarkably effective for denoising, in other words, for the retrieval of an image from noisy observations. They are typically trained on a large dataset made of ground-truth images along with their corrupted versions, which corresponds to a self-supervised learning scheme. 

Denoising networks can then be deployed at each iteration of an optimization framework. This applies in particular to Plug-and-Play (PnP) methods, which are a class of proximal algorithms commonly used to solve~\eqref{eq:minreg}. These solvers alternate between two distinct steps: (a) decrease a data-fidelity loss and (b) correct the current guess by enforcing regularization. This class of algorithms is very flexible. For example, step (a) can represent a proximal operator of the data-fidelity term or a gradient-descent iteration, leading to the alternating-direction method of multipliers (ADMM) and fast iterative shrinkage-thresholding algorithm (FISTA) respectively. 
As discussed in \ref{subsec:conventional regularization}, step (b) has traditionally been a proximal operator of a well-defined regularization term $\mcR$. PnP methods replace this step by the denoising CNN. 

Even though convergence is in general not assured and can only be proven in restricted settings, PnP with a denoising network is extremely flexible and has been used with great success in many applications. For instance, such an approach enables computationally efficient large-scale phase retrieval with state-of-the-art performance~\cite{chang2021large}. 

Likewise, regularization by denoising (RED) \cite{metzler2018prdeep} also exploits inner-loop denoisers to improve the quality of the reconstructions. However, the regularization term is now
\begin{equation}\label{eq:red}
    \mcR(\mathbf{x}) = \mathbf{x}^{\text{H}}(\mathbf{x} - f(\mathbf{x})),
\end{equation}
where $f:\bbC^d \rightarrow \bbC^d$ is the denoiser. This formulation makes the regularization term differentiable, which offers an interesting alternative to PnP. See \cite{kamilov2022plug} for a more in-depth discussion.

\floatstyle{plain}
\restylefloat{figure}
\begin{table*}[!h]
    \centering
    {\small
    \centering
\begin{tabular}{|c|c|c|c|}
\hline
Algorithm & Weak recovery & Perfect recovery & \makecell{Analytic characterization \\ of the performance}
\\
\hline
(Stochastic) Gradient descent & -- & $n \gtrsim \mathcal{O}(d)\,^{(*)}$ & Yes \cite{mignacco2021stochasticity,celentano2021high,sarao2020optimization} \\
PhaseLift \& PhaseCut & -- & $n \gtrsim d \log d$ \cite{candes2013phaselift,waldspurger2015phase} & -- \\
Optimal spectral method $M(\mcT^\star)$ & $n/d \geq 1$ & $n/d \to \infty$ & Yes \cite{luo2019optimal} \\
$M(\mcT^\star)$ + Gradient descent & $n/d \geq 1$ & $n/d \gtrsim 4$ (Fig.~\ref{fig:spectral_methods}) 
& Yes (combination of \cite{luo2019optimal,mignacco2021stochasticity,celentano2021high})\\
AMP & $n/d \geq 1$ & $n/d \gtrsim 2.027$ (Fig.~\ref{fig:amp}) & Yes \cite{maillard2020phase} \\
\hline
\end{tabular}
}
\caption{
    Asymptotic guarantees for the performance of recovery algorithms in the high-dimensional limit, for the Gaussian i.i.d.\ model \ref{model:gaussian_iid} in noiseless complex-valued phase retrieval.
    We add a $\,^{(*)}$ indication when the result requires informed initialization.
}
\label{table:asymptotic_guarantees}

\end{table*}

\subsection{Generative models}
\label{subsec:GANs}

Generative Adversarial Networks (GANs) oppose two networks, a generator $\mcG$ and a discriminator $\mcD$. The goal of the generator is to generate realistic images that follow the distribution of a dataset of natural images. It receives as input a random vector $\bz$ also called \textit{latent variable} and outputs $\bx = \mcG(\bz)$. Then, the task of the discriminator is to choose whether a given image is from the real dataset or some output of the generator. In this adversarial scheme, the two networks are trained together. Throughout training, the generator learns a manifold of realistic images while the discriminator gets better and better at discerning real and fake images.

To use such a generator for phase retrieval, one constrains the image $\bx$ to be an output of the generator $\bx = \mcG(\bz)$ \cite{hand2018phase}. Gradient descent is performed directly on the latent variable $\bz$.
This re-parameterization constrains the search space to the learned manifold of generator outputs. With this point of view, one can not only produce a point estimate of the unknown signal, but also perform a Bayesian sampling of the posterior distribution. This enables the computation of statistical estimates such as the spatial distribution of the variance of the final image \cite{bohra2022bayesian}. 

\subsection{Deep Image Prior}

The machine-learning approaches of Sections \ref{subsec:PnP}--\ref{subsec:GANs} rely on a dataset of images for training. This raises questions about overfitting: by design the final image will resemble a realistic image from the training set while crucial distinctive information may be discarded. Deep Image Prior (DIP) has been proposed to remove any dependence on a training set. Here, one assumes that the estimates are generated from a CNN $\bx = \mcG_\bw(\bz)$, with $\bw$ the weights of the network and $\bz$ a random input. Optimization is then directly performed on the weights $\bw$ of the CNN. DIP has been implemented in several phase-retrieval applications \cite{wang2020phase, bostan2020deep}.

While both the GAN and DIP approaches rely on a generative network, their re-parameterization differ considerably. The former optimizes a low-dimensional latent vector $\bz$, while the second tunes the CNN weights $\bw$. The dimension of $\bw$ is typically larger than the dimension of $\bx$. As a result, DIP can overfit any input $\bx$, even random noise. Nevertheless, it has been observed that it naturally converges at first toward smooth natural images. Regularization comes from the architecture of the CNN itself but requires early stopping to avoid fitting noise. In the end, the exact principle behind DIP regularization remains an open question. 
\label{subsec:DIP}

\subsection{Discussion}

Historically, the first attempts at the application of deep learning targeted a direct inversion of the phase retrieval problem. From the intensity measurements $\by$, they trained a network to output the desired image $\bx$ and applied it to ptychography \cite{kappeler2017ptychnet} or holography \cite{rivenson2018phase}. This strategy corresponds to an usual supervised-learning scheme. However, the algorithms following this strategy are specific to one particular imaging application with a specific forward model, noise level, or camera sampling. This greatly limits their range of applicability. The methods described in Sections \ref{subsec:PnP}--\ref{subsec:DIP} belong to a second generation of algorithms with increased robustness and flexibility. 


All these various strategies demonstrate how deep-learning methods offer a powerful alternative to classical regularization. They open new perspectives for solving the difficult nonlinear equation of phase retrieval. While our understanding is progressing fast, many of these methods remain black boxes that still need to be deciphered. In practice, one also needs to be aware of the notorious instabilities of neural networks \cite{gottschling2020troublesome}, which can sometimes create artefacts (hallucinations) or remove certain features of interest. These issues are particularly troublesome and difficult to detect. The final result may look deceptively realistic (in reason of the regularization inherent with deep learning) even though the algorithm has actually failed. 

From their inception, phase-retrieval algorithms have relied on prior information about the object to reconstruct. Regularization has gone a long way and evolved substantially, from nonnegativity and support constraints to deep learning, but the full potential and limits of the most recent tools at our disposal are not yet clearly established. The majority of the network-based proposed works require a training dataset, which may impede certain applications, 
typically, in medical setups.
DIP and other unsupervised schemes offer an interesting alternative and may pave the way to powerful and controlled regularization strategies based on learning. 
}

\section{Conclusion}

We have described the many facets of phase retrieval, a problem which \new{is central to} a broad range of applications in computational optics. Various experimental configurations and reconstruction algorithms have been proposed to solve the missing-phase problem. To classify such a diversity of applications, we have outlined a unifying framework of many phase-retrieval models. We have also provided a concise overview of the latest phase-retrieval algorithms, to help practitioners choose one that fits their own application. 

Viewed through the prism of machine learning, the links between phase retrieval and neural networks are twofold. First, the parallel between the defining equation~\eqref{eq:pr_definition} and single-layer neural networks has enabled the application of many theoretical machine-learning studies, pushing forward our understanding of this nonlinear equation. Second, the unrivaled performance of neural networks when dealing with images brings strong promises for future applications, but also comes with challenges on how to counteract unstability issues. 

Regarding future developments, the introduction of more randomness in the illumination and/or detection could facilitate phase recovery in experiments, by benefiting from the strong theoretical results obtained for the random setting. However, many realistic phase-retrieval models still remain unexplored from a theoretical viewpoint, calling for more studies. With this document, we hope to promote a fruitful exchange between different disciplines, from applied physics to theoretical computer science. 

\new{\section*{Funding}

J.D., T-.a.P., and M.U. acknowledge funding from European Research Council (ERC) under the European Union’s Horizon 2020 research and innovation programme (Grant Agreement No. 101020573 FunLearn). L.V. acknowledges funding from Swiss National Science Foundation (Grant P400P2\textunderscore199329). S.G. acknowledges funding from European Research Council (ERC) under the European Union’s Horizon 2020 research and innovation programme (Grant Agreement No. 724473 SMARTIES).}

    
    
    
    
    
    

\newrefcontext[]
\printbibliography[keyword = primary, noresetnumbersforlabelprefix, title=References]

@misc{baker2020tramp,
  title         = {Tree-AMP: Compositional Inference with Tree Approximate Message Passing},
  author        = {Antoine Baker and Benjamin Aubin and Florent Krzakala and Lenka Zdeborová},
  year          = {2021},
  eprint        = {2004.01571},
  archiveprefix = {arXiv},
  primaryclass  = {stat.ML}
}

@article{bandeira2014saving,
  title     = {Saving phase: Injectivity and stability for phase retrieval},
  author    = {Bandeira, Afonso S and Cahill, Jameson and Mixon, Dustin G and Nelson, Aaron A},
  journal   = {Applied and Computational Harmonic Analysis},
  volume    = {37},
  number    = {1},
  pages     = {106--125},
  year      = {2014},
  publisher = {Elsevier}
}

@article{candes2013phaselift,
  title     = {Phaselift: Exact and stable signal recovery from magnitude measurements via convex programming},
  author    = {Cand\`es, Emmanuel J and Strohmer, Thomas and Voroninski, Vladislav},
  journal   = {Communications on Pure and Applied Mathematics},
  volume    = {66},
  number    = {8},
  pages     = {1241--1274},
  year      = {2013},
  publisher = {Wiley Online Library}
}

@article{candes2015phase_wf,
  title     = {Phase retrieval via Wirtinger flow: Theory and algorithms},
  author    = {Cand\`es, Emmanuel J and Li, Xiaodong and Soltanolkotabi, Mahdi},
  journal   = {IEEE Transactions on Information Theory},
  volume    = {61},
  number    = {4},
  pages     = {1985--2007},
  year      = {2015},
  publisher = {IEEE}
}

@inproceedings{celentano2020estimation,
  title        = {The estimation error of general first order methods},
  author       = {Celentano, Michael and Montanari, Andrea and Wu, Yuchen},
  booktitle    = {Conference on Learning Theory},
  pages        = {1078--1141},
  year         = {2020},
  organization = {PMLR}
}

@article{chang2021large,
  title     = {Large-scale phase retrieval},
  author    = {Chang, Xuyang and Bian, Liheng and Zhang, Jun},
  journal   = {eLight},
  volume    = {1},
  number    = {1},
  pages     = {1--12},
  year      = {2021},
  publisher = {Springer}
}

@article{fannjiang2020numerics,
  title     = {The numerics of phase retrieval},
  author    = {Fannjiang, Albert and Strohmer, Thomas},
  journal   = {Acta Numerica},
  volume    = {29},
  pages     = {125--228},
  year      = {2020},
  publisher = {Cambridge University Press}
}

@article{fienup1987phase,
  title   = {Phase retrieval and image reconstruction for astronomy},
  author  = {Fienup, James R and Dainty, Chris},
  journal = {Image Recovery: Theory and Application},
  volume  = {231},
  pages   = {275},
  year    = {1987}
}

@book{goodman2005introduction,
  title     = {Introduction to Fourier optics},
  publisher = {Roberts and Company Publishers},
  year      = {2005},
  author    = {Goodman, Joseph W}
}

@article{guizar2008phase,
  title     = {Phase retrieval with transverse translation diversity: a nonlinear optimization approach},
  author    = {Guizar-Sicairos, Manuel and Fienup, James R},
  journal   = {Optics Express},
  volume    = {16},
  number    = {10},
  pages     = {7264--7278},
  year      = {2008},
  publisher = {Optical Society of America}
}

@inproceedings{hand2018phase,
  title     = {Phase retrieval under a generative prior},
  author    = {Hand, Paul and Leong, Oscar and Voroninski, Vladislav},
  booktitle = {Proceedings of the 32nd International Conference on Neural Information Processing Systems},
  pages     = {9154--9164},
  year      = {2018}
}

@article{luo2019optimal,
  title     = {Optimal spectral initialization for signal recovery with applications to phase retrieval},
  author    = {Luo, Wangyu and Alghamdi, Wael and Lu, Yue M},
  journal   = {IEEE Transactions on Signal Processing},
  volume    = {67},
  number    = {9},
  pages     = {2347--2356},
  year      = {2019},
  publisher = {IEEE}
}

@article{maiden2009improved,
  title     = {An improved ptychographical phase retrieval algorithm for diffractive imaging},
  author    = {Maiden, Andrew M and Rodenburg, John M},
  journal   = {Ultramicroscopy},
  volume    = {109},
  number    = {10},
  pages     = {1256--1262},
  year      = {2009},
  publisher = {Elsevier}
}

@article{maillard2020phase,
  title   = {Phase retrieval in high dimensions: Statistical and computational phase transitions},
  author  = {Maillard, Antoine and Loureiro, Bruno and Krzakala, Florent and Zdeborov{\'a}, Lenka},
  journal = {Advances in Neural Information Processing Systems},
  volume  = {33},
  year    = {2020}
}

@inproceedings{metzler2017coherent,
  title        = {Coherent inverse scattering via transmission matrices: Efficient phase retrieval algorithms and a public dataset},
  author       = {Metzler, Christopher A and Sharma, Manoj K and Nagesh, Sudarshan and Baraniuk, Richard G and Cossairt, Oliver and Veeraraghavan, Ashok},
  booktitle    = {2017 IEEE International Conference on Computational Photography (ICCP)},
  pages        = {1--16},
  year         = {2017},
  organization = {IEEE}
}

@inproceedings{metzler2018prdeep,
  title        = {prDeep: Robust phase retrieval with a flexible deep network},
  author       = {Metzler, Christopher and Schniter, Phillip and Veeraraghavan, Ashok and others},
  booktitle    = {International Conference on Machine Learning},
  pages        = {3501--3510},
  year         = {2018},
  organization = {PMLR}
}

@article{miao1999extending,
  title     = {Extending the methodology of X-ray crystallography to allow imaging of micrometre-sized non-crystalline specimens},
  author    = {Miao, Jianwei and Charalambous, Pambos and Kirz, Janos and Sayre, David},
  journal   = {Nature},
  volume    = {400},
  number    = {6742},
  pages     = {342--344},
  year      = {1999},
  publisher = {Nature Publishing Group}
}

@article{mignacco2021stochasticity,
  title     = {Stochasticity helps to navigate rough landscapes: comparing gradient-descent-based algorithms in the phase retrieval problem},
  author    = {Mignacco, Francesca and Urbani, Pierfrancesco and Zdeborov{\'a}, Lenka},
  journal   = {Machine Learning: Science and Technology},
  year      = {2021},
  publisher = {IOP Publishing}
}

@article{mondelli2019fundamental,
  title   = {Fundamental Limits of Weak Recovery with Applications to Phase Retrieval},
  author  = {Mondelli, Marco and Montanari, Andrea},
  journal = {Foundations of Computational Mathematics},
  volume  = {19},
  number  = {3},
  year    = {2019}
}

@article{paxman1992joint,
  title     = {Joint estimation of object and aberrations by using phase diversity},
  author    = {Paxman, Richard G and Schulz, Timothy J and Fienup, James R},
  journal   = {JOSA A},
  volume    = {9},
  number    = {7},
  pages     = {1072--1085},
  year      = {1992},
  publisher = {Optical Society of America}
}

@article{rodenburg2019ptychography,
  title     = {Ptychography},
  author    = {Rodenburg, John and Maiden, Andrew},
  journal   = {Springer Handbook of Microscopy},
  pages     = {819--904},
  year      = {2019},
  publisher = {Springer}
}

@article{sayre1952some,
  title     = {Some implications of a theorem due to Shannon},
  author    = {Sayre, David},
  journal   = {Acta Crystallographica},
  volume    = {5},
  number    = {6},
  pages     = {843--843},
  year      = {1952},
  publisher = {International Union of Crystallography}
}

@inproceedings{schniter2016vector,
  title        = {Vector approximate message passing for the generalized linear model},
  author       = {Schniter, Philip and Rangan, Sundeep and Fletcher, Alyson K},
  booktitle    = {2016 50th Asilomar Conference on Signals, Systems and Computers},
  pages        = {1525--1529},
  year         = {2016},
  organization = {IEEE}
}

@article{shechtman2015phase,
  title     = {Phase retrieval with application to optical imaging: a contemporary overview},
  author    = {Shechtman, Yoav and Eldar, Yonina C and Cohen, Oren and Chapman, Henry Nicholas and Miao, Jianwei and Segev, Mordechai},
  journal   = {IEEE signal processing magazine},
  volume    = {32},
  number    = {3},
  pages     = {87--109},
  year      = {2015},
  publisher = {IEEE}
}

@article{yeh2015experimental,
  title     = {Experimental robustness of Fourier ptychography phase retrieval algorithms},
  author    = {Yeh, Li-Hao and Dong, Jonathan and Zhong, Jingshan and Tian, Lei and Chen, Michael and Tang, Gongguo and Soltanolkotabi, Mahdi and Waller, Laura},
  journal   = {Optics Express},
  volume    = {23},
  number    = {26},
  pages     = {33214--33240},
  year      = {2015},
  publisher = {Optical Society of America}
}

@article{zheng2013wide,
  title     = {Wide-field, high-resolution Fourier ptychographic microscopy},
  author    = {Zheng, Guoan and Horstmeyer, Roarke and Yang, Changhuei},
  journal   = {Nature Photonics},
  volume    = {7},
  number    = {9},
  pages     = {739--745},
  year      = {2013},
  publisher = {Nature Publishing Group}
}

@article{kamilov2022plug,
  title={Plug-and-play methods for integrating physical and learned models in computational imaging},
  author={Kamilov, Ulugbek S and Bouman, Charles A and Buzzard, Gregery T and Wohlberg, Brendt},
  journal={arXiv preprint arXiv:2203.17061},
  year={2022}
}

@article{bostan2020deep,
  title={Deep phase decoder: self-calibrating phase microscopy with an untrained deep neural network},
  author={Bostan, Emrah and Heckel, Reinhard and Chen, Michael and Kellman, Michael and Waller, Laura},
  journal={Optica},
  volume={7},
  number={6},
  pages={559--562},
  year={2020},
  publisher={Optica Publishing Group}
}

@inproceedings{kappeler2017ptychnet,
  title={Ptychnet: CNN based Fourier ptychography},
  author={Kappeler, Armin and Ghosh, Sushobhan and Holloway, Jason and Cossairt, Oliver and Katsaggelos, Aggelos},
  booktitle={2017 IEEE International Conference on Image Processing (ICIP)},
  pages={1712--1716},
  year={2017},
  organization={IEEE}
}

@article{rivenson2018phase,
  title={Phase recovery and holographic image reconstruction using deep learning in neural networks},
  author={Rivenson, Yair and Zhang, Yibo and G{\"u}nayd{\i}n, Harun and Teng, Da and Ozcan, Aydogan},
  journal={Light: Science \& Applications},
  volume={7},
  number={2},
  pages={17141--17141},
  year={2018},
  publisher={Nature Publishing Group}
}

@article{bohra2022bayesian,
  title={Bayesian Inversion for Nonlinear Imaging Models using Deep Generative Priors},
  author={Bohra, Pakshal and Pham, Thanh-an and Dong, Jonathan and Unser, Michael},
  journal={arXiv preprint arXiv:2203.10078},
  year={2022}
}

@article{sarao2020optimization,
  title={Optimization and generalization of shallow neural networks with quadratic activation functions},
  author={Sarao Mannelli, Stefano and Vanden-Eijnden, Eric and Zdeborov{\'a}, Lenka},
  journal={Advances in Neural Information Processing Systems},
  volume={33},
  pages={13445--13455},
  year={2020}
}

@article{metzler2020deep,
  title={Deep-inverse correlography: towards real-time high-resolution non-line-of-sight imaging},
  author={Metzler, Christopher A and Heide, Felix and Rangarajan, Prasana and Balaji, Muralidhar Madabhushi and Viswanath, Aparna and Veeraraghavan, Ashok and Baraniuk, Richard G},
  journal={Optica},
  volume={7},
  number={1},
  pages={63--71},
  year={2020},
  publisher={Optical Society of America}
}

@article{candes2015phase,
  title     = {Phase retrieval from coded diffraction patterns},
  author    = {Cand\`es, Emmanuel J and Li, Xiaodong and Soltanolkotabi, Mahdi},
  journal   = {Applied and Computational Harmonic Analysis},
  volume    = {39},
  number    = {2},
  pages     = {277--299},
  year      = {2015},
  publisher = {Elsevier}
}

@article{celentano2021high,
  title   = {The high-dimensional asymptotics of first order methods with random data},
  author  = {Celentano, Michael and Cheng, Chen and Montanari, Andrea},
  journal = {arXiv preprint arXiv:2112.07572},
  year    = {2021}
}

@article{chang2018hybrid,
  title     = {Hybrid optical-electronic convolutional neural networks with optimized diffractive optics for image classification},
  author    = {Chang, Julie and Sitzmann, Vincent and Dun, Xiong and Heidrich, Wolfgang and Wetzstein, Gordon},
  journal   = {Scientific Reports},
  volume    = {8},
  number    = {1},
  pages     = {1--10},
  year      = {2018},
  publisher = {Nature Publishing Group}
}

@article{conca2015algebraic,
  title     = {An algebraic characterization of injectivity in phase retrieval},
  author    = {Conca, Aldo and Edidin, Dan and Hering, Milena and Vinzant, Cynthia},
  journal   = {Applied and Computational Harmonic Analysis},
  volume    = {38},
  number    = {2},
  pages     = {346--356},
  year      = {2015},
  publisher = {Elsevier}
}

@article{gerchberg1972practical,
  title   = {A practical algorithm for the determination of phase from image and diffraction plane pictures},
  author  = {Gerchberg, Ralph W},
  journal = {Optik},
  volume  = {35},
  pages   = {237--246},
  year    = {1972}
}

@article{goldstein2018phasemax,
  title     = {Phasemax: Convex phase retrieval via basis pursuit},
  author    = {Goldstein, Tom and Studer, Christoph},
  journal   = {IEEE Transactions on Information Theory},
  volume    = {64},
  number    = {4},
  pages     = {2675--2689},
  year      = {2018},
  publisher = {IEEE}
}

@article{gottschling2020troublesome,
  title   = {The troublesome kernel: why deep learning for inverse problems is typically unstable},
  author  = {Gottschling, Nina M and Antun, Vegard and Adcock, Ben and Hansen, Anders C},
  journal = {arXiv preprint arXiv:2001.01258},
  year    = {2020}
}

@article{katz2014non,
  title     = {Non-invasive single-shot imaging through scattering layers and around corners via speckle correlations},
  author    = {Katz, Ori and Heidmann, Pierre and Fink, Mathias and Gigan, Sylvain},
  journal   = {Nature photonics},
  volume    = {8},
  number    = {10},
  pages     = {784--790},
  year      = {2014},
  publisher = {Nature Publishing Group}
}

@inproceedings{maillard2021construction,
  title        = {Construction of optimal spectral methods in phase retrieval},
  author       = {Maillard, Antoine and Krzakala, Florent and Lu, Yue M and Zdeborov{\'a}, Lenka},
  booktitle    = {Mathematical and Scientific Machine Learning},
  volume       = {145},
  pages        = {1-28},
  year         = {2021},
  organization = {PMLR}
}

@book{mezard2009information,
  title     = {Information, Physics, and Computation},
  author    = {M\'ezard, Marc and Montanari, Andrea},
  year      = {2009},
  publisher = {Oxford University Press}
}

@article{valzania2021accelerating,
  title     = {Accelerating ptychographic reconstructions using spectral initializations},
  author    = {Valzania, Lorenzo and Dong, Jonathan and Gigan, Sylvain},
  journal   = {Optics Letters},
  volume    = {46},
  number    = {6},
  pages     = {1357--1360},
  year      = {2021},
  publisher = {Optical Society of America}
}

@inproceedings{vinzant2015small,
  title        = {A small frame and a certificate of its injectivity},
  author       = {Vinzant, Cynthia},
  booktitle    = {2015 International Conference on Sampling Theory and Applications (SampTA)},
  pages        = {197--200},
  year         = {2015},
  organization = {IEEE}
}

@article{waldspurger2015phase,
  title     = {Phase recovery, maxcut and complex semidefinite programming},
  author    = {Waldspurger, Ir\`ene and d’Aspremont, Alexandre and Mallat, St{\'e}phane},
  journal   = {Mathematical Programming},
  volume    = {149},
  number    = {1},
  pages     = {47--81},
  year      = {2015},
  publisher = {Springer}
}

@article{wang2020phase,
  title     = {Phase imaging with an untrained neural network},
  author    = {Wang, Fei and Bian, Yaoming and Wang, Haichao and Lyu, Meng and Pedrini, Giancarlo and Osten, Wolfgang and Barbastathis, George and Situ, Guohai},
  journal   = {Light: Science \& Applications},
  volume    = {9},
  number    = {1},
  pages     = {1--7},
  year      = {2020},
  publisher = {Nature Publishing Group}
}

@article{wu2018lensless,
  title     = {Lensless digital holographic microscopy and its applications in biomedicine and environmental monitoring},
  author    = {Wu, Yichen and Ozcan, Aydogan},
  journal   = {Methods},
  volume    = {136},
  pages     = {4--16},
  year      = {2018},
  publisher = {Elsevier}
}

@inproceedings{yurtsever2017sketchy,
  title        = {Sketchy decisions: Convex low-rank matrix optimization with optimal storage},
  author       = {Yurtsever, Alp and Udell, Madeleine and Tropp, Joel and Cevher, Volkan},
  booktitle    = {Artificial intelligence and statistics},
  pages        = {1188--1196},
  year         = {2017},
  organization = {PMLR}
}

@article{zhang20173d,
  title     = {3D computer-generated holography by non-convex optimization},
  author    = {Zhang, Jingzhao and P{\'e}gard, Nicolas and Zhong, Jingshan and Adesnik, Hillel and Waller, Laura},
  journal   = {Optica},
  volume    = {4},
  number    = {10},
  pages     = {1306--1313},
  year      = {2017},
  publisher = {Optical Society of America}
}
\end{document}